\documentclass[doublecol]{epl2} 

\title{ Discrete elastic model for stretching-induced  flagellar polymorphs}

\author{H. Wada\inst{1,2} \and R. R. Netz\inst{1}}

\institute{                    
 \inst{1} Physics Department, Technical University Munich, 85748 Garching, Germany\\
 \inst{2} Yukawa Institute for Theoretical Physics, Kyoto Universiy, Kyoto, 606-8502, Japan
}

\pacs{87.16.Ka}{Filaments, microtubules, their networks, and supermolecular assemblies}
\pacs{87.15.-v}{Biomolecules; structure and physical properties}
\pacs{64.70.Nd}{Structural transitions in nanoscale materials}

\newcommand{\bra}{\langle}
\newcommand{\ket}{\rangle}
\newcommand{\vecOm}{\boldsymbol{\Omega}}

\newcommand{\vecxi}{\boldsymbol{\xi}}

\abstract{
Force-induced reversible transformations between coiled and normal 
polymorphs of bacterial flagella have been observed in recent
optical-tweezer experiment.
We introduce a discrete elastic rod model with two competing  
helical states governed by a fluctuating spin-like variable 
that represents the underlying conformational states of flagellin monomers.
Using hybrid Brownian dynamics Monte-Carlo simulations, we show 
that a helix undergoes shape transitions dominated by
domain wall nucleation and motion
in response to externally applied uniaxial tension.
A scaling argument for the critical force is presented in
good agreement with experimental and simulation results. 
Stretching rate-dependent elasticity including a buckling instability 
are found, also consistent with the experiment.}

\begin{document}

\maketitle

{\bf Introduction -- }
Many motile bacteria achieve directed propulsion in a viscous environment 
by rotating flagellar filaments~\cite{berg-anderson}. 
A number of common bacteria, such as {\it Escherichia coli} or {\it Salmonella},
are externally flagellated, where each flagellum consists
of  a helically-shaped  elastic filament that is attached to
a rotary motor embedded in the rigid cell wall via a flexible hook.
The flagellar filament is a large homogeneous assembly of a single protein,
flagellin, but transforms between several helical shapes of different
pitch, radius and handedness in 
response to  environmental or mechanical stimuli such as change of
temperature, pH, or externally applied forces and torques~\cite{darnton-turner-berg}.
For example, {\it Salmonella} swim by rotating their left-handed helical
filaments counter-clockwise, forming a co-ordinated bundle
of filaments~\cite{kim-powers,reinchert-stark}.
The reverse of the motor rotational direction gives rise to 
a tumbling motion of the cell body, often (but not always)  accompanied by
changes in filament chirality to right-handed~\cite{darnton-turner-berg,macnab-ornston}.
By regulating the relative durations of  swimming and tumbling modes, bacteria
are able to  migrate through a stimulus gradient in the  favorable direction.

The flagellar filament is composed of eleven protofilaments 
that each consist of a stack of flagellin monomers and that are wrapped
around each other~\cite{namba}.
As first proposed by Asakura~\cite{asakura} and later proved by 
crystallography~\cite{yamashita}, 
a flagellin monomer can take two conformations, L-type and R-type, that
are slightly different in intrinsic twist and length. Each
protofilament can be assumed to consist entirely of either  L or R monomers, however,
the complete filament is thought to contain variable numbers of 
L and R-type  protofilaments.
Depending on how many protofilaments are of the L or R type, 
the flagellum  takes different polymorphic helical states,
thereby minimizing the elastic strain energy.
A simple geometric model developed by Calladine in line with 
this idea~\cite{calladine} explained the observed spectrum of 
flagellar polymorphic forms.
A coarse-grained continuum rod model has also been developed
recently~\cite{srigiriraju-powers}.

The Asakura-Calladine model has revealed the design principle of
bacterial flagellar filament. Recently,
dynamics of flagellar polymorphic transitions have moved into the focus.
Turner {\it et al.} have reported an intricate sequence of transformations 
of fluorescently labelled {\it E. coli} flagella during a single 
run-to-tumble mode switching~\cite{turner}. Clearly, understanding
elastic energy barriers  between polymorphic forms
is crucial for the understanding of switching dynamics.
Several studies, both experimental and theoretical, have 
addressed this issue~\cite{darnton-turner-berg,hotani,washizu,coombs}.
Hotani reported cyclic chirality transformations of {\it Salmonella} flagella
tethered at one end to a glass surface and subjected to an external 
fluid flow~\cite{hotani}.
Above a certain critical flow velocity (a few $\mu$m/s), chirality flips occur 
at the tethered end and nucleated chirality domains grow and 
propagate steadily down the filament.
The magnitudes of pulling force or  mechanical torque necessary for the onset of 
the normal-to-semi-coiled transition
were estimated as $15$ pN and $10^{-18}$ N m respectively, 
based on a slender-body hydrodynamics analysis of the experimental geometry.
Coombs and Goldstein~\cite{coombs} theoretically analysed and rationalized
Hotani's observation based on a bistable helix model
introduced earlier by Goldstein and co-workers~\cite{goldstein}.
Using an optical tweezer setup, Darnton and Berg have recently measured the 
force-extension curves of {\it Salmonella} flagella filaments~\cite{darnton-berg}.
Mechanical stretching induces a reversible polymorphic transformation between 
coiled and normal state at a threshold force around 3-5 pN, characterized
by large hysteresis during pulling and relaxing.
The measured force is shown to be strongly rate-dependent, suggesting
that the observed polymorphic transformation is kinetically controlled.
Stretching experiments of {\it Salmonella} flagella have also been
performed using atomic force microscopy by Cluzel \etal~\cite{cluzel}.
Since atomistic simulations are impossible with present computer power
due to the huge size of a flagellum and the large time scales involved, 
coarse-graining is imperative. In one recent
simulation study, a flagellin monomer is represented by 15 point masses with 
interactions parameterized based on the known protein crystal structure\cite{Schulten}.
Using an implicit solvent model, different helical structures are observed 
depending on rotational direction and in qualitative accord with experiments.
But still, the numerical simulation effort only allows to simulate the model
flagellum for about 30 $\mu$s which is short compared to intrinsic time scales
of polymorphic dynamics\cite{Schulten}. It is therefore desirable to take
the coarse-graining to an even higher level.

In this paper, we introduce an elastic helix model that is 
coupled to an internal discrete variable that locally describes polymorphic states.
In the present formulation we consider an Ising-like spin variable 
that switches between two states of different helix pitch and radius.
We introduce a hybrid Brownian-dynamics Monte-Carlo simulation and
show that our model reproduces well the observed force-extension
relationship of flagellar filaments accompanying polymorphic 
transformations~\cite{darnton-berg}. 
As in experiments, the force-extension curves are strongly 
stretching-rate dependent.
We also give simple scaling arguments for the critical switching force, in 
good agreement with experimental and our numerical data. 

{\bf Model -- }
Within linear elasticity theory, the bending and twisting 
energy for an isotropic helical rod of contour length $L$ 
parameterized by arclength $s$ is given by
\begin{equation}
E_{el} = \frac{1}{2}\int_0^L ds \left[ A \Omega_1^2+A (\Omega_2-\Omega_2^0)^2
 + C (\Omega_3-\Omega_{3}^0)^2 \right],
 \label{energy}
\end{equation}
where $A$ and $C$ stand, respectively, for the bending and 
twisting modulus~\cite{love}.
$\vecOm=(\Omega_1,\Omega_2,\Omega_3)$ is the 
strain rate vector; the curvature $\kappa$ satisfies
$\kappa^2=\Omega_1^2+\Omega_2^2$ and $\Omega_3$ is the twist density.
The ground-state shape of a filament is  
specified by the intrinsic curvature $\kappa_0$ and torsion $\tau_0$,
related to the radius $R$ and pitch $P$ via 
$\kappa_0=4\pi^2 R/(P^2+4\pi^2R^2)$
and $\tau_0=2\pi P/(P^2+4\pi^2R^2)$.
Alternatively, one can specify 
the pitch angle $\psi$ through $\tan\psi=2\pi R/P$, and the contour
length of a single helical turn, $\ell=(P^2+4\pi^2R^2)^{1/2}$.
The relation between these two descriptions is given by
$\kappa_0=(2\pi/\ell)\sin\psi$ and 
$\tau_0=(2\pi/\ell)\cos\psi$.
Experimentally, the so-called 
``coiled'' state of a filament is the  equilibrium shape at  pH 4 and
 $3^{\circ}$ C (see the phase diagram in ref.~\cite{darnton-berg}).
We in particular  study  the polymorphic transition
from the coiled to the so-called  ''normal'' state.
For the coiled and normal forms of {\it Salmonella}, which we 
respectively label as 1 and 2, $\ell$ and
$\psi$ were measured as 
$(\ell_1,\ell_2)=(3.4\,\mu\tx{m},2.5\,\mu\tx{m})$ and 
$(\psi_1,\psi_2)=(76.7^{\circ},31.3^{\circ})$~\cite{darnton-berg}.
In the present study, we for simplicity set the length $\ell$ for 
both forms equal  and choose
pitch angles $\psi_1=73.3^{\circ}$ and $\psi_2=29.7^{\circ}$,
close to the experimental values.
The intrinsic curvature and torsion for each helical state follow as
$\kappa_{0 m}=(2\pi/\ell)\sin\psi_m$ 
and $\tau_{0 m}=(2\pi/\ell)\cos\psi_m$, where $m=1,2$.

In the simulations
we consider a chain of $N+1$ connected spheres of diameter $a$.
Each bead is specified by its position ${\bf r}_j$ and rotation
angle about the local tangent $\phi_j$, from which the discrete
strain rate ${\bf \Omega}_j$ is calculated.
Analogous to the helix-coil model for biopolymers~\cite{zimm-bragg},
each sphere carries a discrete variable $\sigma_j$, where 
$\sigma_j=1$ specifies the coiled form and
$\sigma_j=-1$ the normal form.
To account for cooperativity, induced by elastic strain
at the interface between two polymorphic states,
a short-range interaction $J$ between neighbouring ``spin'' variables
$\sigma_j$ and  $\sigma_{j+1}$ is included.
The total  energy functional of the system (apart from a stretching
contribution introduced later) reads  in discretized form
\begin{eqnarray}
&& E[{\bf \Omega},\sigma] =
  - J\sum_{\bra i,j\ket}\sigma_i\sigma_j-ha \sum_{j=1}^N\sigma_j  \\
&& +   \frac{a}{2}\sum_{j=1}^N\left[ A \Omega_{1j}^2
  + A \left(\Omega_{2j}-\Omega^0_{2j}  \right)^2
 + C \left(\Omega_{3j}-\Omega^0_{3j}   \right)^2 \right], \nonumber 
 \label{energy2}
\end{eqnarray} 
where $\bra i,j\ket$ implies summation over  nearest neighbors 
and the $\sigma$-dependent intrinsic strain rates are 
$\Omega_{2j}^0= \sigma_j  (\kappa_{01}-\kappa_{02})/2
+(\kappa_{01}+\kappa_{02})/2$, and
$\Omega_{3j}^0= \sigma_j (\tau_{01}-\tau_{02})/2
+(\tau_{01}+\tau_{02})/2$.
The chemical potential bias per length, $h$, accounts for 
a preference of one state over the other,
and depends on environmental conditions such as 
solution pH or salinity.

\begin{figure}
\onefigure[width=0.80\linewidth]{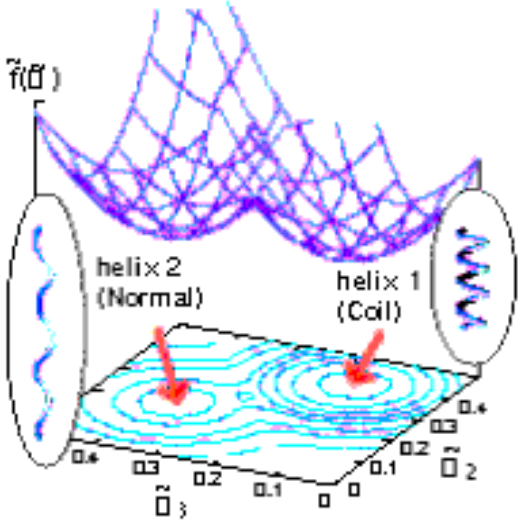}
\caption{Plot of the rescaled effective elastic free energy
$\tilde{f}=fa/k_BT$, eq.~(\ref{effective_f2}), 
as function of curvature and twist strain rates $\Omega_2, \Omega_3$
(with $\Omega_1=0$).
The two local minima correspond to the normal and coiled form
as indicated.
Parameters values are $A/(ak_BT)=1844$, $C/A=0.8$,
$J/k_BT=10$, and $h/k_BT=7.7$.}
\label{fig1}
\end{figure}

{\bf Effective potential --} To gain insight into the coupling between
helix elasticity and the polymorphic degrees of freedom, we formally
define an effective elastic energy density $f({\vecOm})$ 
by integrating out the variables $\sigma$ as 
\begin{equation}
 \exp\left[-\frac{a}{k_BT}\sum_{j=1}^N f({\vecOm}_j)\right]  
	= \sum_{\{\sigma\}}
	e^{-E[{\bf \Omega},\sigma]/k_BT}
 \label{partition}
\end{equation}
where $k_B T$ denotes thermal energy.
For uniform helical shape, {\it i.e.}, $j$-independent strain rates,
$\Omega_{2j}=\Omega_2$ and $\Omega_{3j}=\Omega_3$, the summation over 
$\sigma$ can be  performed using transfer matrix method.
Assuming periodic boundary condition $\sigma_1=\sigma_N$, and
discarding unimportant constants, we obtain
\begin{widetext}
\begin{equation}
\label{effective_f2}
 f(\vecOm) = \frac{A}{2}\Omega_1^2+\frac{A}{2}\left[\Omega_2
	-\frac{\kappa_{01}+\kappa_{02}}{2}\right]^2
	+\frac{C}{2}\left[\Omega_3-\frac{\tau_{01}+\tau_{02}}{2}\right]^2
	-k_BT\ln\left[e^{J/k_BT}\cosh\Lambda + 
	\sqrt{e^{2J/k_BT}\sinh^2\Lambda+e^{-2J/k_BT}}\right],
\end{equation}
\end{widetext}
\begin{floatequation}
\mbox{\textit{see eq.$~(\ref{effective_f2})$}}
\end{floatequation}
where we defined 
$k_BT\Lambda({\vecOm})= ha+
\frac{Aa}{2}(\kappa_{01}-\kappa_{02})
[\Omega_2-\frac{1}{2}(\kappa_{01}+\kappa_{02})] 
+\frac{Ca}{2}(\tau_{01}-\tau_{02})
[\Omega_3-\frac{1}{2}(\tau_{01}+\tau_{02})]$ which
acts as a strain-dependent external field. The effective
elastic energy  $f({\vecOm})$ is plotted in fig.~\ref{fig1}
as a function of the two relevant strain rates
for one specific parameter set.
The potential $f$ is a  double-well 
with two  minima corresponding to the normal
and the coiled form.
Previous theoretic treatments used a very similar
shape of the elastic free energy as a starting point\cite{coombs}.

{\bf Simulation details --}
In our hybrid simulations, the energy functional in addition
includes a  stretching
contribution that ensures  connectivity of the spheres, 
$E_{st}=K/2\sum_{j=1}^N(|{\bf r}_{j+1}-{\bf r}_j|-a)^2$.
The local elastic translational force ${\bf F}_j$ and torque 
about the local tangent $T_j$ acting on each sphere are calculated
using the variational method described previously~\cite{chirico,wada-netz}, 
leading to the coupled Langevin equations:
\begin{equation}
 \partial_t{\bf r}_i = \mu_0{\bf F}_i+\vecxi_i(t),
 \quad
 \mbox{and}
 \quad
 \partial_t\phi_i = \mu_r T_i+\Xi_i(t),
 \label{langevin}
\end{equation}
where $\partial_t$ is the time derivative.
Neglecting hydrodynamic effects, we take the mobility matrix 
to be diagonal and use the Stokes translational and rotational mobilities of a sphere 
$\mu_0=1/(3\pi\eta a)$ and $\mu_r=1/(\pi\eta a^3)$, 
where $\eta$ is the solvent viscosity.
The vectorial random forcings $\vecxi(t)$ and $\Xi(t)$ model
the coupling to a heat bath and obey the fluctuation-dissipation
relations.
Standard Monte-Carlo Metropolis methods are applied to update $\sigma$.
In each Brownian dynamics time step, 20 Monte-Carlo steps 
are carried out, meaning that equilibration of the $\sigma$ degrees of freedom
is much faster than filament shape relaxations.

For the simulations we discretize the Langevin equations (\ref{langevin})
with time step $\Delta$ and rescale all lengths,
times and energies;  the rescaled time step
$\tilde{\Delta}=\Delta k_BT\mu_0/a^2$ is
for sufficient numerical accuracy chosen in the range
$\tilde{\Delta}=10^{-4}$-$10^{-5}$.
The stretching modulus is set to $K/k_BTa^2=10^4$ 
which keeps  bond length fluctuations negligibly small.
Observables are calculated every $10^3$-$10^4$ steps, total 
simulation times are in the order of $10^{6}$-$10^{8}$ steps.
One filament end is fixed at the origin, and the other end
(initially being at its equilibrium position, {\it i.e.}, 
$z =L\cos\psi_1$) is moved along the $\hat{\bf z}$-axis 
(identical to the helix axis).
We set the maximum extension $z=0.8L$, the total stretching 
distance is thus $D=(0.8-\cos\psi_1)L$.
We fix the number of beads $L/a=N=60$ and the arclength of one helical 
turn $\ell=15a$, the
pitch angle $\psi$ thus uniquely determines the
helical shape of the filament.
We set the bend/twist ratio $\Gamma=C/A=0.8$ with 
$\tilde{A}=A/(ak_BT)=1844$.
The bend persistence length is thus $L_p/L\simeq 31$, 
comparable to the experimental conditions
$L_p/L\simeq 40-120$, obtained from measured values of the bending 
stiffness $A\simeq 3.5$ pN$\cdot \mu$m$^2$ and typical filament length
of 7.6-19.5 $\mu$m~\cite{darnton-berg}.
Although the magnitude of $J$ is experimentally unknown, 
we assume $J/k_BT=10$
sufficiently larger than thermal energy in accord with
the observation that thermally
assisted nucleation of polymorphic transitions is absent in the stress-free 
state~\cite{coombs,goldstein}.
The rescaled stretching speed, $\tilde{V}=Va/(\mu_0k_BT)$, is changed
in the range $\tilde{V}=0.01$-0.2, to study systematically 
the rate-dependent elasticity.
Assuming $a\simeq 126$ nm (giving $L=aN\simeq 7.6$ $\mu$m for $N=60$
as in the experiment~\cite{darnton-berg}), we obtain 
$0.3-3.3$ $\mu$m/s comparable to the experimental 
$V\simeq 0.4$ $\mu$m/s~\cite{darnton-berg}.
Our filament is thus thicker than a real bacterial flagellar filament
by roughly a factor of 5, which however is not serious as
the filament thickness enters hydrodynamic drag coefficients
only logarithmically.
The pulling end of the filament is free
to rotate, {\it i.e.}, no external torque is applied, while
the tethered end is not allowed to rotate, suitable to 
the experiment condition.

\begin{figure}
\onefigure[width=0.99\linewidth]{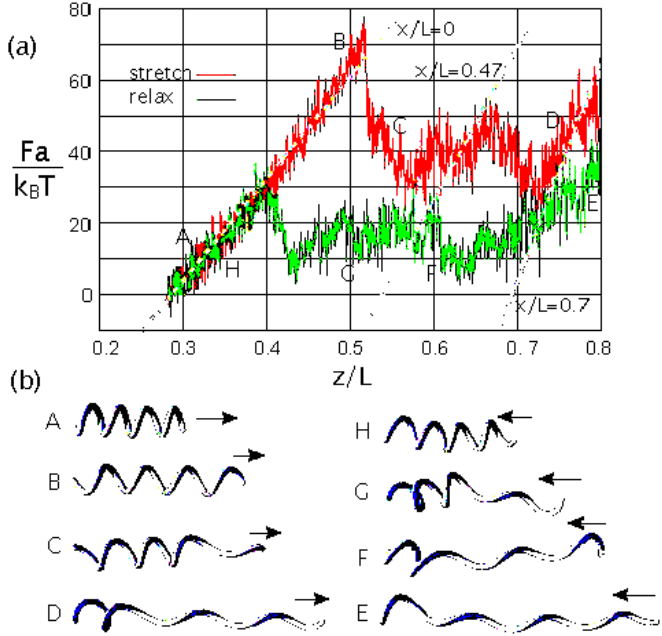}
\caption{(a) Typical force-extension curve obtained for $N=60$, $\tilde{A}=1844$,
$C/A=0.8$, $J/k_BT=10$ and $\tilde{h}=ha/k_BT=7.7$ 
at a stretching speed $\tilde{V}=0.0124$.
Red line is the stretching, the green one is the relaxing curve.
The broken lines follow from the mixed Hookean-spring theory  eq.~(\ref{f-mixedspring})
for $x/L=0, 0.47$ and 0.7, from left to right.
(b) Corresponding snapshots of the filament at points specified
in fig.~\ref{fig2} (a).}
\label{fig2}
\end{figure}

\begin{figure}
\onefigure[width=0.65\linewidth]{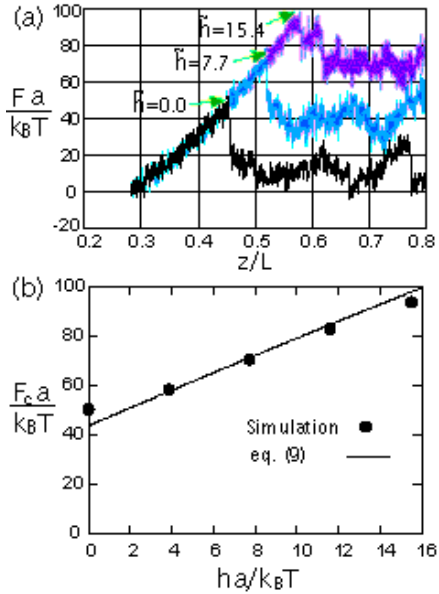}
\caption{(a) Stretching force curves for varying
chemical potential bias $\tilde{h}=0.0, 7.7, 15.4$, 
obtained at $\tilde{V}=0.0124$.
All other parameters are the same as in fig.~\ref{fig2} (a). 
(b) Threshold force, $\tilde{F}_c$, obtained from fig.~\ref{fig3} (a),
plotted as a function of $\tilde{h}$. Broken line is the analytic
result, eq.~(\ref{f_c}), with $x\simeq 1.32\ell$ as the best fit.}
\label{fig3}
\end{figure}

{\bf Force-induced shape transformations --}
In fig.~\ref{fig2}, we show one typical force-extension curve
obtained at pulling rate $\tilde{V}=0.0124$ (the slowest considered), 
with several snapshots of the filament 
at the points specified in  (a).
The initial filament shape is coiled,
according to the experimental conditions~\cite{darnton-berg}. 
As the extension $z$ increases, the helix 
deforms uniformly like a linear spring.
Neglecting  thermal effects, 
the restoring force $F$ of a deformed
helix is on the linear-elasticity level given by~\cite{wada-netz}
\begin{equation}
 F = AC\frac{(\kappa_{01}\cos\psi-\tau_{01}\sin\psi)
	(A\kappa_{01}\sin\psi+C\tau_{01}\cos\psi)}{\sin\psi
	(A\sin^2\psi+C\cos^2\psi)^2},
 \label{fxcurve-spring}
\end{equation}
with $z/L=\cos\psi$. 
For small forces, Hooke's law,
$F \simeq K_{1,L}(z-L\cos\psi_1)$, holds, where 
$K_{m,L}=\partial{F}/\partial{z}|_{z=L\cos\psi_m}
=(4\pi^2/\ell^2)(L\sin^2\psi_m)^{-1}(\cos^2\psi_m/A+
\sin^2\psi_m/C)^{-1}$ with  $m=1,2$, is the linear spring constant~\cite{wada-netz,love}.
The spring constant is inversely proportional to the arclength of the helix.
At a certain critical force ($\tilde{F}\simeq 70$ in fig.~\ref{fig2} (a)),
the uniform helical shape becomes metastable, and 
a shape transformation occurs from the moving end of the
filament via nucleation of the normal form, see the snapshot C in
fig.~\ref{fig2} (b).
The filament separates into two domains of
different pitch angles $\psi_1$ and $\psi_2$, and
the tension $F$ suddenly drops down to much smaller value; 
compare B \& C in fig.~\ref{fig2}.
After the first transition, subsequent more gradual
shape transformations at 
increasing  extension $z$ are observed at lower forces.
Approximating the filament as two serial Hookean springs of arclength 
$L-x$ and $x$, we obtain a modified Hooke's law
\begin{equation}
 F = \frac{K_{1,L-x}K_{2,x}}{K_{1,L-x}+K_{2,x}}
	\left[z-(L-x)\cos\psi_1-x\cos\psi_2\right].
 \label{f-mixedspring}
\end{equation}
In fig.~\ref{fig2} (a), eq.~(\ref{f-mixedspring}) 
is compared to the numerical data
assuming helix compositions of $x/L=$ 0, 0.47 and 0.7 (broken lines) and
captures reasonably well the response of the mixed filament.
The main point here is that the helical transformation proceeds in subsequent steps.

A normal helical section with pitch angle $\psi_2$ appears upon
stretching via nucleation.
Assuming the helix as a linear spring, 
the filament elastic energy  (relative to its stress-free state) 
at force $F$ is 
$E \simeq \frac{1}{2}F^2/K_{1,L}$.
Once a normal section of arclength $x$ appears, the tensile force
drops down to $F'$ at constant  extension $z$.
We note that in the simulations we almost always 
observe nucleation accompanied by two domain walls.
The energy of the mixed state at the  transition thus is
$E'\simeq \frac{1}{2}(1/K_{1,L-x}+1/K_{2,x})F'^2+2hx+4J$.
Nucleation proceeds only when $E'<  E$.
Using $F'=(1/K_{1,L-x}+1/K_{2,x})^{-1} (F/K_{1,L}-\gamma)$,
where $\gamma=x(\cos\psi_2-\cos\psi_1)$, and assuming
$K_{1,L-x}/K_{1,L}\simeq 1$ valid for $x/L\ll 1$, we 
finally obtain the threshold force $F_c$ as a function of $x$
and $h$:
\begin{equation}
 F_c \simeq -\gamma K_2+\sqrt{(K_1+K_2)[K_2\gamma^2+
	4(hx+2J)]},
 \label{f_c}
\end{equation}
where we have used the shorthand notation $K_{1,L-x}=K_1$ and $K_{2,x}=K_2$.

We perform stretching simulations for five different values of
the chemical potential bias, $\tilde{h}$. 
All other parameters are the same as those in fig.~\ref{fig2} (a).
The force-extension curves for $\tilde{h}=0, 7.7, 15.4$ are shown 
in fig.~\ref{fig3} (a), from which
the critical force $F_c$ is obtained as a function of $h$, which
is plotted in fig.~\ref{fig3} (b).
Since the nucleation domain size $x$ is not determined 
by our arguments, we fit $x\simeq 19.8a\simeq 1.32\ell$
in the comparison of
eq.~(\ref{f_c}) to the numerical data in fig.~\ref{fig3} (b),
which is also consistent with the nucleating
domain size extracted from the snapshots
B and C in fig.~\ref{fig2} (b).
The agreement between the numerical data and eq.~(\ref{f_c}) is 
excellent, verifying the mixed
Hookean spring scaling model. 

To demonstrate the experimental relevance of our results, we  put
 experimental values into eq.~(\ref{f_c}), namely
$\psi_1=76.7^{\circ}$, $\psi_2=31.3^{\circ}$, 
$\ell_1=3.4$ $\mu$m, $\ell_2=2.5$ $\mu$m, $A\simeq 3.5$ pN $\mu$m$^2$, 
$L\simeq 7.6$ $\mu$m~\cite{darnton-berg} and 
$x\simeq 1.32 \ell_2 \simeq 3.35$ $\mu$m from fig.~\ref{fig3} (b). 
The experimental value of the energy difference between 
the two polymorphs, $h$, is unknown, and 
we obtain $F_c\simeq 3-5$ pN for the range $h \simeq 0.15-0.76$ pN 
$\simeq 0.04-0.20k_BT$/nm.
The experimentally determined critical force $F_c$ 
for the coiled-to-normal transition
ranges over 3-5 pN~\cite{darnton-berg}, quite consistent with our results.
The direct (and independent) experimental measurement of $h$ might be hard, but
atomistic simulations~\cite{kitao} 
may  in the future check our prediction of $h$.

The contraction process involves discontinuous reverse 
polymorphic transformations characterized by sudden increases of
the tensile force; see fig.~\ref{fig2} (b). 
The reverse transitions are in general observed at much lower forces
and at different extensions, resulting in a large hysteresis.

\begin{figure}
\onefigure[width=0.99\linewidth]{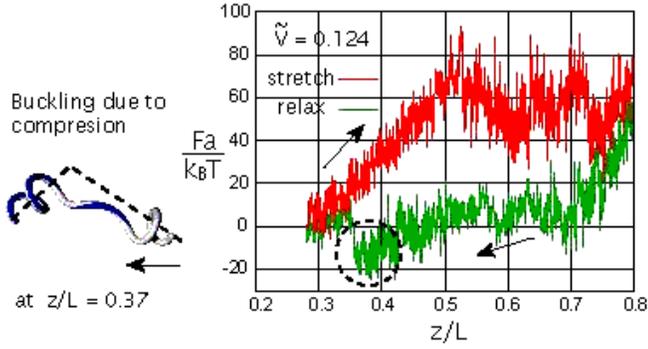}
\caption{Force-extension curve at velocity $\tilde{V}=0.124$, 
ten times faster than fig.~\ref{fig2} (a). Shown is one stretching
(red line) contraction (green) cycle. 
The snapshot of the buckling state (left)  is obtained at stretching
$z/L=0.37$ upon contraction and yields negative forces
(highlighted by broken circle).
All other parameters are the same as used in fig.~\ref{fig2} (a).}
\label{fig4}
\end{figure}

{\bf Kinetic effects --}
In fig.~\ref{fig4}, we show a force-extension curve for the fastest velocity
studied by us,
$\tilde{V}=0.124$, with
all other parameters as in fig.~\ref{fig2} (a).
Notable is the negative force in the contraction 
process at $\tilde{V}=0.124$.
This is due to a buckling of the helix via compression;
a  buckled filament configuration is shown to the left.
A buckling-induced negative force has also been observed 
in the experiment~\cite{darnton-berg}.

\begin{figure}
\onefigure[width=0.75\linewidth]{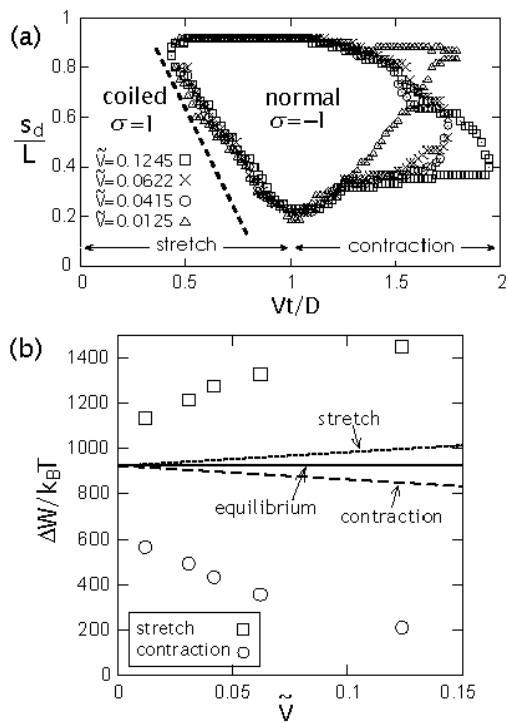}
\caption{
(a) Trajectories of the rescaled domain-wall positions $s_d/L$ 
as a function of the reduced time $Vt/D$ for varying velocities
$\tilde{V}$.
The slope of the broken line is $(\cos\psi_2-\cos\psi_1)^{-1}$
from eq.~(\ref{c_v}).
(b) Mechanical work done between $z_1=0.3L$ and $z_2=0.8L$, 
$\Delta W/k_BT$,  as a function of $\tilde{V}$ 
in the stretching (square) and contraction (circle) processes. 
Each data point is the average over three statistically
independent runs. (Error bars are smaller than the symbols.)
The solid line indicates the equilibrium work, and the broken
lines are the friction work for stretching and contraction processes (see
text). Parameters are the same as in fig.~\ref{fig2} (a),
unless otherwise stated.}
\label{fig5}
\end{figure}

The domain wall between  normal and coiled sections
moves in response to the pulling of the filament end.
We plot in fig.~\ref{fig5} (a) the domain wall positions $s_d$
as a function of 
the reduced time $V t/D$, where $D$ is the stretching distance defined
earlier and thus $D/V$ gives the total stretching time.
The domain boundary position  is obtained from the discrete
$\sigma$-profile along the filament. As noted earlier,
we generally observe the occurrence of two domain walls.
The trajectories in fig.~\ref{fig5} (a) are strongly asymmetric between 
the stretch and contraction processes.
On the stretching side, $0<Vt/D \leq1$, trajectories
for different velocity $V$ collapse.
The domain-wall speed $\partial{s_d}/\partial{t}$ upon stretching
approximately follows from a simple geometric argument as
\begin{equation}
 \frac{\partial{s_d}}{\partial{t}} 
 = \frac{V}{\cos\psi_2-\cos\psi_1}.
 \label{c_v}
\end{equation}
To see this, note that  the lengths of the two helices
are $s_d$ and $L-s_d$.
Neglecting helix deformations  from its intrinsic
shape, the end-point position follows as
$z\simeq s_d\cos\psi_1+(L-s_d)\cos\psi_2$, 
which upon inversion yields Eq.(\ref{c_v})
where   $V=\partial{z}/\partial{t}$ is the imposed stretching speed.
In the reduced units in fig.~\ref{fig5} (a), this gives the line
of slope $(\cos\psi_2-\cos\psi_1)^{-1}$ independent of $V$, 
which is shown as a broken line, in
reasonable agreement with the numerical data at early times.
Deviations at later stages are connected to nonlinear filament deformations.

To quantify rate-dependent dissipation effects,
we calculate the work done between $z_1=0.3L$ and $z_2=0.8L$,
$\Delta W=\int_{z_1}^{z_2} dz F$, in both the stretching and 
contraction processes, which is shown in
fig.~\ref{fig5} (b) as a function of $V$.
The difference between the two works, which is a measure of dissipation, 
becomes larger as $V$ increases. 
Especially for small $V$, $\Delta W$
is strongly influenced by the stretching rate $V$,
consistent with the experimental observations~\cite{darnton-berg}.
One trivial contribution to dissipation is 
due to solvent friction. 
Assuming the filament to be homogeneously stretched
and on the free-draining level (as appropriate for our simulations), the 
friction work is
$\Delta W_F\simeq \pi\eta a N (z_2-z_1) V$,
which is shown in fig.~\ref{fig5} (b) as broken lines.
As one can see, the solvent friction does not account for the entire dissipation,
the difference is due to elastic dissipation within the filament.
In the quasi-static limit $V\rightarrow 0$, $\Delta W$ for stretching 
and contraction should become equal.
Although a direct numerical check of this is difficult
because of prohibitively long simulation times, 
a simple scaling estimate is helpful:
Assuming that the helix takes a perfect normal form (all $\sigma=-1$) 
at high stretching  $z=0.8L$ without any elastic deformation, 
and a perfect coiled form with all $\sigma=1$
at $z=0.3L$, the total elastic energy change is simply 
$\Delta E=2hL$ which determines the work $\Delta W$ in the quasistatic limit
at $V=0$. In rescaled units, this leads to
$\Delta W/k_BT\simeq 2\tilde{h}N=924$ for $\tilde{h}=7.7$,
which in fig.~\ref{fig5} (b) is indicated with a solid horizontal line and is flanked
by the stretching and contraction data as expected.
In the large $V$ region, i.e. for $\tilde{V}>0.06$ in fig.~\ref{fig5} (b),
which probably is beyond experimentally accessible speeds, 
the force response seems less affected by the stretching
rate $V$ and the apparent rate-dependence is largely due
to solvent friction.

{\bf Summary --} In this letter, 
we have introduced  a bistable helical filament model 
that accounts for different elastic monomeric  states by 
a discrete Ising-like spin variable along the arclength.
Employing hybrid  Brownian dynamics Monte-Carlo simulations,
we have studied the kinetic effects of uniaxial stretching on the helical shape.
The main results are:
(i) The helical filament undergoes 
 reversible polymorphic transformations in response to an externally 
applied uniaxial tension, consistent with the optical-tweezer
experiments of {\it Salmonella} flagellar filaments.
(ii) Experimental force-extension curves including
sawtooth pattern and large hysteresis are qualitatively reproduced.
(iii) A simple analytic argument based on a mixed Hookean spring
model for the critical force of the
polymorph induction provides good agreement with numerical and 
experimental data.
(iv) Force-extension curves are strongly  pulling rate-dependent, 
in particular a buckling-induced negative force is observed, 
all consistent with the experimental observations.

Generalization of our model to include all possible polymorphic flagellar states
is straightforward but includes many at present unknown parameters.
Applying our model to torque-induced chirality transformations of
flagellar filaments, directly relevant to 
bacterial motility in vivo, is in progress~\cite{wada-netz-prepri}.

Financial support from Research Abroad Program of the Japan Society for
the Promotion of Science (JSPS) and
the German Science Foundation (DFG, SPP1164 and SFB 486) is acknowledged. 


\end{document}